\documentclass[floatfix,showpacs,superscriptaddress,nofootinbib,twocolumn,showkeys,prl]{revtex4-2}

\usepackage[utf8]{inputenc}

\usepackage{graphicx}
\usepackage{xcolor}
\usepackage{amsmath}
\usepackage{hyperref}
\usepackage{siunitx}
\usepackage{amssymb}
\usepackage{multirow,array}
\usepackage{lipsum}
\usepackage{enumitem}

\def\fluc#1{ \left\langle #1 \right\rangle }

\newcommand{\prlsection}[1]{\noindent \textbf{\emph{#1}} -- }

\usepackage{soul}
\setstcolor{red}

\begin{document}

\title{
Heavy-Ion Collisions as Probes of Nuclear Structure
}

\author{Nicolas Mir\'{o} Fortier}
\affiliation{Department of Physics, McGill University, Montr\'{e}al QC H3A\,2T8, Canada}

\author{Sangyong Jeon}
\affiliation{Department of Physics, McGill University, Montr\'{e}al QC H3A\,2T8, Canada}
\author{Charles Gale}
\affiliation{Department of Physics, McGill University, Montr\'{e}al QC H3A\,2T8, Canada}

\date{\today}

\begin{abstract}
 In this work, we perform a 
model-to-data comparison for U+U and Au+Au collisions performed at 
RHIC 
at $\sqrt{s_{\rm NN}}$ = $\qty{193}{\giga\electronvolt}$ and 
$\qty{200}{\giga\electronvolt}$, using a multistage
framework. 
Model calculations for various configurations
of $^{238}$U and $^{197}$Au
are used to compare with measurements of 
$\rho(v_n\{2\}^2,\fluc{p_T})$, the elliptic-flow-momentum correlator, for the first time. It is found that momentum-flow correlations measured in high-energy nuclear collisions, 
combined with the flow harmonics themselves, are excellent probes of 
nuclear deformation and of nuclear structure in general.
\end{abstract}

\maketitle

\prlsection{Introduction} Experiments performed at heavy-ion collider facilities have been successful in 
confirming the existence of the quark-gluon plasma (QGP), 
an extreme state of strongly interacting matter predicted by QCD \cite{Busza_2018}. 
Theoretical analyses of the experimental 
evidence suggests that QGP expands and cools like a fluid, 
opening the door for modelling using viscous 
relativistic hydrodynamics~\cite{Gale:2013da,Pasechnik_2017}. Computational heavy-ion
collision models have been in use for the better part of 2 decades
and have been providing grounds for testing physical descriptions of QCD 
under extreme conditions.

Anisotropic flow has been a key observable marker of collective behavior in QGP~\cite{Gale:2013da,Jeon_2015,Kolb:1999it}, and has as such 
been the subject of a large number of analyses since the advent of heavy-ion collision simulations.
The Fourier components $v_n$ of the azimuthal momentum distribution of particles produced in collision events
can be used to decompose anisotropic flow into components: elliptic flow $v_2$ measures its ellipticity,
triangular flow $v_3$ measures its triangularity, and so on~\cite{Bilandzic:2012wva}. 
This flow and its components are directly related
to the properties of the produced QGP droplet,
not only through its transport coefficients~\cite{Heinz_2013} but also through its geometric properties:
a more elliptically-shaped QGP will lead to more elliptic flow in the final state. Ellipticity has two
sources in heavy-ion collisions. The first is centrality:
when two spherically-symmetric nuclei collide at non-zero
impact parameter $b$, they produce an elliptic overlap region which in turn generates elliptic flow.
The second is deformity:
when two prolate (or rugby-ball-shaped) nuclei collide,
the overlap region can take different shapes, even at the same impact parameter $b$,
depending on the relative orientation of the nuclei.
This relative orientation also affects the energy density,
especially in ultra-central collisions. It therefore creates a strong correlation between 
the energy density and the shape of the interaction region.

This combination of ellipticity and energy leads to tangible effects in non-flow observables as well.
For instance, this set of differing anisotropies in central collisions provides interesting 
grounds for measuring the impact of overlap region size and shape on mean transverse momentum
$\fluc{p_T}$. Indeed, given that collisions belonging to the same centrality bins have similar 
amounts of total energy deposited in the interaction region, nuclear deformation acts as a further geometric 
modulating tool: we can isolate the effects of initial state anisotropies stemming from deformation 
in those collisions.
Furthermore, given the wide spectrum of possible 
overlap configurations in central collisions, we expect noticeable effects on $\fluc{p_T}$ 
among events in central bins. 
Since vastly different events (in terms of anisotropy)
populate 
individual centrality bins, unique effects of deformation can be measured in final-state observables, 
such as an anti-correlation between elliptic flow $v_2$ and mean transverse momentum $\fluc{p_T}$.
In contrast, this correlation has been observed to be positive through all centralities of spherical
nuclei collisions~\cite{Schenke_transverse,Aad_2019}.

Using our state-of-the-art model consisting of IP-Glasma~\cite{GELIS_CGC,Iancu_2004,Gale:2012rq,Gale:2013da,Schenke_2020,heffernan2023bayesian,McDonald_2017,heffernan2023earlytimes,mcdonald202331d}, relativistic viscous hydrodynamics via MUSIC~\cite{MUSIC_Schenke,Ryu_2015}, iS3D~\cite{mcnelis2020particlization}, and SMASH~\cite{Weil_2016} 
particlization and final states, we will show that nuclear structure plays a key role in the interpretation of a wide variety of observables.
In particular, we will demonstrate the effectiveness of correlation 
measurements in isolating the impact of nuclear deformation on heavy-ion collisions, suggesting future analyses of experimental data should incorporate our findings.

\prlsection{Nuclear Structure} 
The modelling of nuclei in the study of heavy-ion collisions
has evolved over the past two decades, together with the increased sophistication of the initial states used in multi-stage approaches.  
Indeed, accurate representations of the nuclear shape 
deformity have become key to reproducing larger collections of related observables 
simultaneously. 

The standard Woods-Saxon distribution~\cite{PhysRev.95.577} 
can accommodate nuclear deformity as follows
\begin{gather}
    \rho(r,\theta,\phi) = \frac{\rho_0}{1 + \exp \left(\frac{r - R(\theta,\phi)}{a}\right)} \\
    R(\theta, \phi) = R_0\left( 1 + \sum_{l = 2}^{l_{max}} \sum_{m=-l}^{l} \beta_l^m Y_l^m(\theta,\phi) \right) \label{eq:ws}
\end{gather}
where $\rho_0$ is the normal nuclear density adjusted to account for the atomic mass number,
$R_0$ is the unmodified nuclear radius,
and $a$ is nuclear skin depth.
The shape modifications are provided by the functional form $R(\theta,\phi)$, which introduces 
nuclear radius dependence on the polar and azimuthal angles via deformation parameters 
$\beta_l^m = \beta_{l}^{-m}$ and
spherical harmonics $Y_l^m(\theta,\phi)$.

\begin{table}
    \caption{\label{tab:params} Woods-Saxon parameters used in the sampling of $^{238}$U and
    $^{197}$Au nuclei. The parameters are from Ref.~\cite{osti_6477756} for Prev U and Spher Au and Ref.~\cite{Ryssens_2023} for New U and Def Au.
    }
    \begin{ruledtabular}
        \begin{tabular}{cccccccc}
         & $R_0$ (fm) & a (fm) & $\beta_2^0$ & $\beta_2^2$ & $\beta_4^0$ & $\beta_4^2$ & $\beta_4^4$
         \\
        \hline
        New U & 7.068 & 0.538 & 0.247 & 0 & 0.081 & 0 & 0 \\
        Prev U & 6.874 & 0.556 & 0.2802 & 0 & -0.0035 & 0 & 0 \\
        Spher Au & 6.37 & 0.535 & 0 & 0 & 0 & 0 & 0 \\
        Def Au & 6.62 & 0.519 & 0.098 & 0.076 & -0.025 & -0.018 & -0.018\\
        \end{tabular}
        \end{ruledtabular}
	\label{table:WS_parameters}
\end{table}

\begin{figure}[t]
    \includegraphics[width=\linewidth]{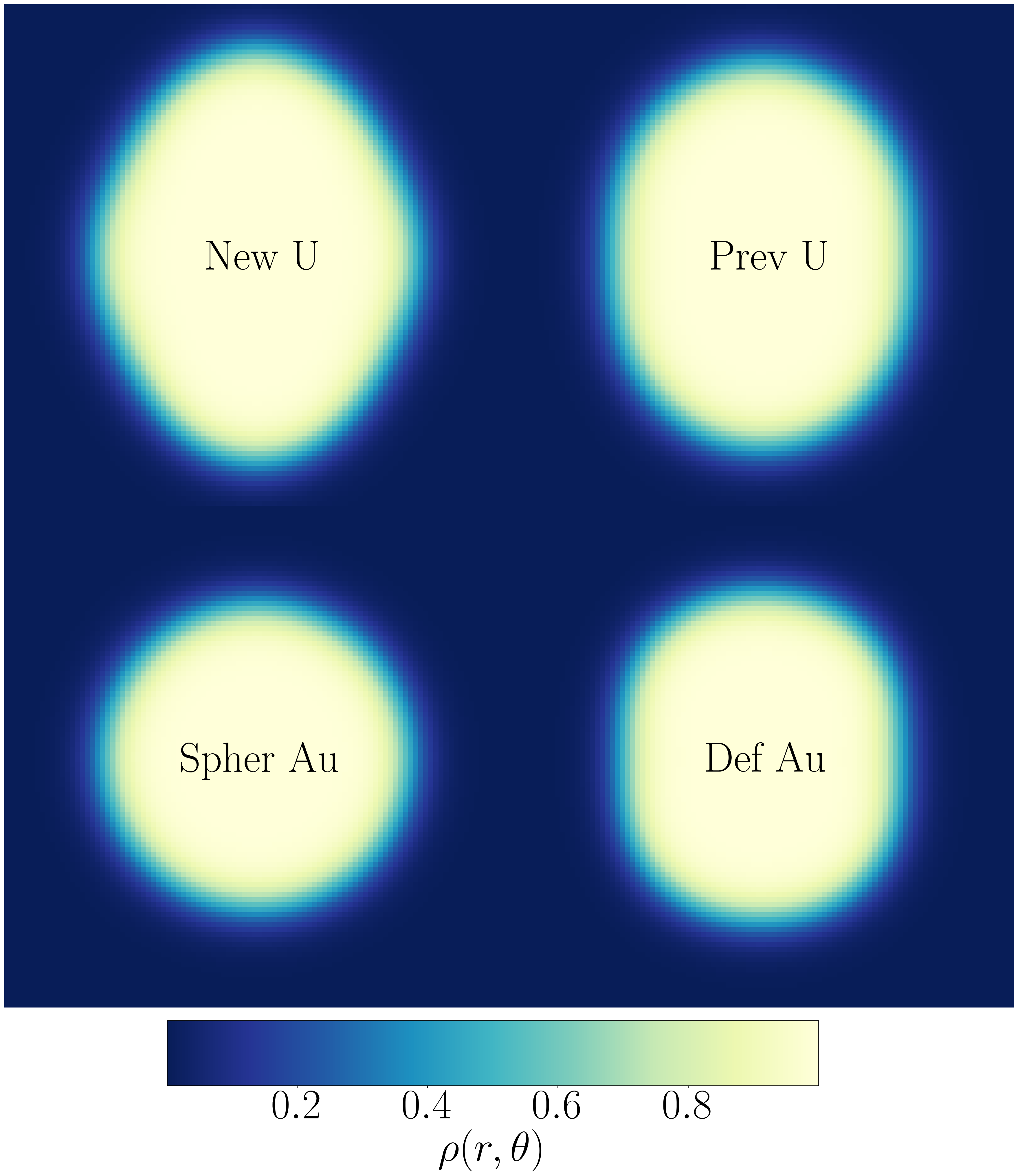}
    \caption{\label{fig:params} Comparing all Woods-Saxon distributions from Tab.~\ref{tab:params} with $\rho_0 = 1$ for comparison's sake. The size ordering
    between $^{238}$U and $^{197}$Au is apparent. Def Au cross-section taken at $\phi=\pi/2$.
    All other cross-sections do not depend on $\phi$.}
\end{figure}

The two nuclei of interest in this study are $^{238}$U and $^{197}$Au. They are relatively close in mass number,
but their shapes exhibit differences. Those are made apparent in Fig.~\ref{fig:params}, 
where the Woods-Saxon distributions $\rho(r,\theta)$ of the 4 parametrizations listed in
Table~\ref{tab:params} are plotted in 2D.
It is important to note that, given the $\phi$
dependence of the Def Au parametrization (non-zero $\beta_l^m$ parameters where $m \neq 0$
), the cross-section varies considerably with $\phi$: at $\phi = 0$, the 
radius function appears quasi-circular, while at $\phi = \pi/2$, the value used in Fig.~\ref{fig:params}, the nucleus
appears appreciably deformed. These different parametrizations were developed
using different methods. The Prev U and Spher Au 
parametrizations come from electron scattering 
experiments~\cite{osti_6477756}, and have been used in simulations many 
times over~\cite{Schenke_2020,Schenke_2014,BBUUIPG,Bally_2023,Magdy:2022cvt}. The New U and Def Au 
parametrizations, on the other hand, stem from a new analysis using 
state-of-the-art Skyrme-Hartree-Fock-Bogoliubov (HFB) calculations~\cite{Ryssens_2023}. 
These parametrizations differ considerably from their counterparts, 
motivating their consideration in our analysis. Therefore, a pair of 
`older' parametrizations originate from low-energy experiments, while the more recent
ones come from an extension of low-energy experimental data~\cite{Pritychenko_2016} to Woods-Saxon parameters using novel methods, namely Skyrme-HFB calculations.

Recent modelling paradigms based on Nuclear Density Functional Theory (NDFT) have proven 
useful~\cite{doi:10.1080/23746149.2020.1740061,Xu_2023} in providing a more
first-principle-based nuclear structure model. 
These models -- and other microscopic approaches -- can provide physical features which are not present in
Woods-Saxon modelling, such as nucleon-nucleon correlations.
However, the
gains provided by these types of models are outweighed by their computational costs.
Fortunately, for this study, the only relevant degree of freedom required is that of nuclear deformation, which can be efficiently parameterized by Eq.(\ref{eq:ws}) above.

Because nuclear shapes have a tangible effect on final-state observables, it is possible to select some that are especially sensitive to those nuclear deformations. 
Furthermore,
as will be clear in the next section, these effects are more pronounced  in the $0-10\%$ centrality window.
Indeed, in more peripheral collisions, it becomes more difficult
to isolate the effects of 
the underlying structure of atomic nuclei.
Therefore, 
the calculations in this study concentrate on the  region 
up to $\sim 25\%$ centrality.  Our centrality binning techniques are outlined
in more detail in our companion paper~\cite{fortier2023comparisons}.

\prlsection{Flow Coefficients and Correlations} In the introduction, we underlined 
the importance of understanding and differentiating between dominant sources 
of flow anisotropies as well as their relationship with final-state observables.
We now present results showing that our model not only matches experimental data across systems, but allows for 
discriminating between closely related nuclear parametrizations and, therefore, selection of the most 
appropriate parametrizations based on a multi-observable analysis. Our model was calibrated a single time against U+U $N_{\textrm{CH}}$ v. centrality curves~\cite{Adamczyk_2015} and 
hydrodynamic parameters were chosen according to recent studies~\cite{Schenke_2020,heffernan2023bayesian}. 
This calibration entails setting the initial energy in the fluid dynamical simulation to a value such that the charged
particle yield  is reproduced. For a more detailed discussion regarding calibration and observable definitions, please refer to our companion paper~\cite{fortier2023comparisons}.
\begin{figure}
    \includegraphics[width=\linewidth]{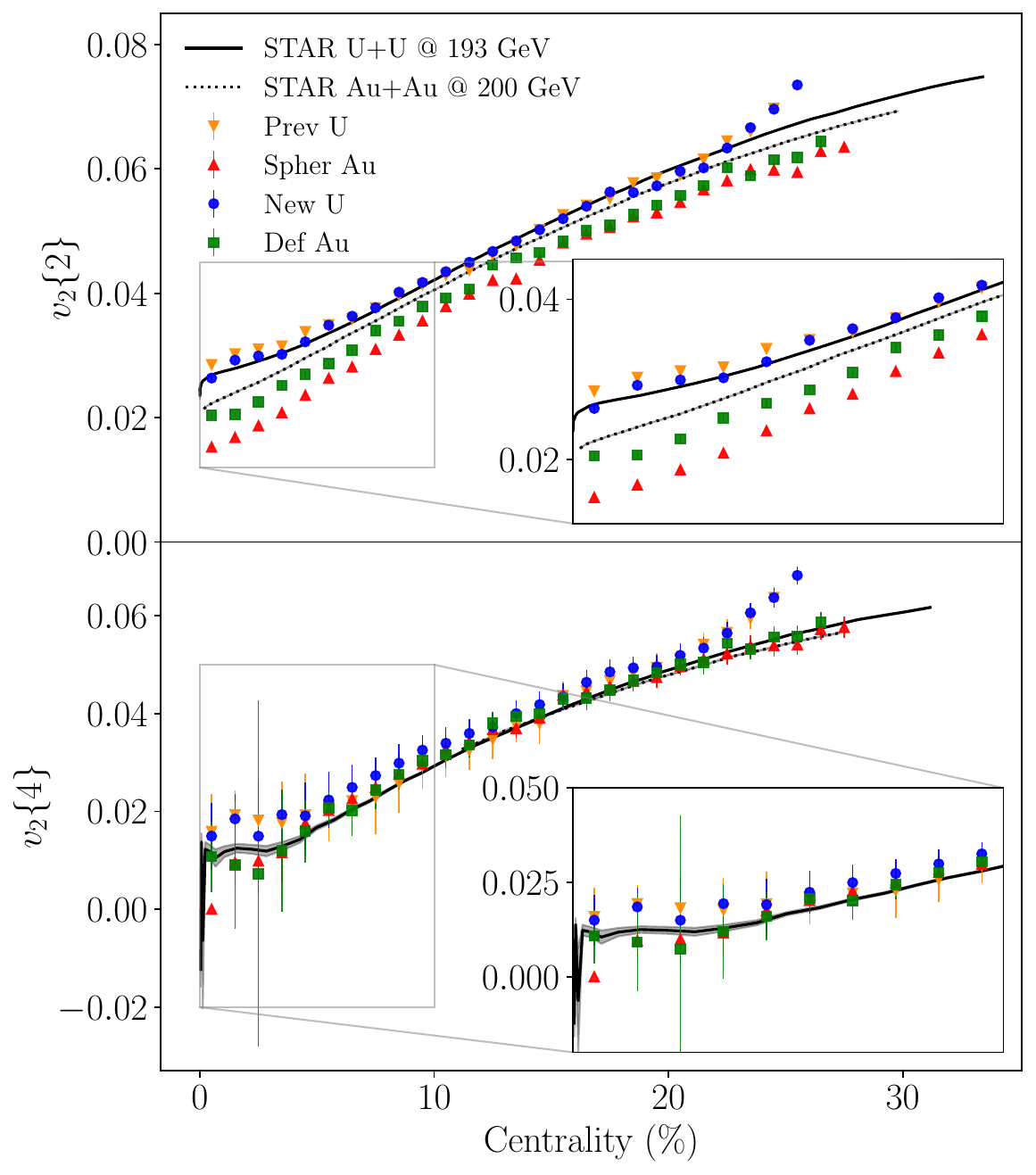}
    \caption{\label{fig:v2} Shown are 2- and 4-particle cumulants of elliptic flow ($v_2\{2\}$ and $v_2\{4\}$) as functions
    of centrality, compared to results for $\qty{193}{\giga\electronvolt}$ U+U and 
    $\qty{200}{\giga\electronvolt}$ Au+Au collisions measured by STAR~\cite{Adamczyk_2015, Abelev_2009}.
    The shaded bands represent statistical and systematic errors. Insets show the 0-10\% centrality region. Experimental errors represented by shaded bands surrounding result 
    curves. 4-particle cumulants have larger errors as they implicitly include the 
    errors of the 2-particle cumulants, amongst other quantities, as described in~\cite{Bilandzic:2012wva}.}
\end{figure}

Fig.~\ref{fig:v2} shows $v_2\{2\}$ and $v_2\{4\}$, the 2- and 4-particle cumulants of 
elliptic flow, as functions of collision centrality. The 4-particle cumulants
$v_2\{4\}$ seems to be unaffected by changes in collision systems and underlying 
nuclear structure. Indeed, across our two systems and four parametrizations, no 
discernible or meaningful difference exists between the curves. Furthermore, the
experimental curves essentially overlap across the entire centrality range, with 
only slight differences in peripheral collisions making themselves seen at the $20\%$ mark.
This makes sense given that 2-particle contributions are subtracted from 4-particle contributions
when calculating $v_2\{4\}$~\cite{fortier2023comparisons,Bilandzic_2011}. As such,
as we move from 2-particle cumulants towards cumulants of 4-, 6- and 8-particles, we 
are probing for effects of increasing range and inherence. That is, 
given that contributions from prior cumulants are subtracted from 4-, 6- and 8-
particle cumulants, the effects of collision geometry which are visible in the 2-particle 
cumulant of the elliptic flow are inevitably washed out of subsequent cumulants by 
construction; because we are interested in the \textit{true} 4-particle correlation, we must
account for correlations which are already found in the 2-particle cumulant.
This 2-particle cumulant $v_2\{2\}$, on the other hand, exhibits a clear dependence on the 
underlying nuclear structure, as is evidenced by the gaps between 
the curves at centralities smaller than $10\%$. We can see that experimental
data prefers the New U and Def Au parametrizations. Our analysis is therefore consistent
with the most recent and advanced of structure calculations~\cite{Ryssens_2023}. The difference between 
the two U parametrizations is slight but noticeable. Indeed, the fairly small change in 
$\beta_2^0$ and $\beta_4^0$ between the parametrizations leads to better agreement with
the experimental data. The gap between the two Au configurations,
on the other hand,
entails that Au is truly deformed, and attempting to match these experimental data 
with a spherically symmetric nuclear parametrization may not be appropriate.
Moreover, $v_2\{2\}$ is better 
equipped at separating between the two distinct systems and collision energies, 
especially in central collisions. Indeed, the experimental U+U curve shows a noticeable
change of slope around $1{-}5\%$ centrality, which then quickly drops to values
similar to those of Au+Au at the $0{-}1\%$ bin. This wrinkle is due to the considerable deformation of $^{238}$U. In central
collisions ($b \approx \qty{0}{\femto\meter}$), $^{238}$U provides two distinct full-overlap 
configurations: one where the short-axes of the nuclei are aligned with the 
beam-pipe and their long-axes are aligned with one another (called body-body) 
collisions, and another where the nuclei's long-axes are aligned with the beam-pipe 
(called tip-tip).
Body-body collisions have a larger overlap area resulting in
less dense nuclear matter at overlap, leading to slightly less particles generated in 
the final state than in tip-tip collisions which, in contrast, generate higher matter densities and 
smaller overlap areas.
Body-body collisions also have a naturally eccentric nature, given their profiles,
as is evidenced in Fig.~\ref{fig:params}. The $5\%$ centrality bin is where body-body
collisions start to dominate. The distribution then dips at $0-1\%$ because ultra-central
collisions are mostly of the tip-tip type.

\begin{figure}
    \includegraphics[width=\linewidth]{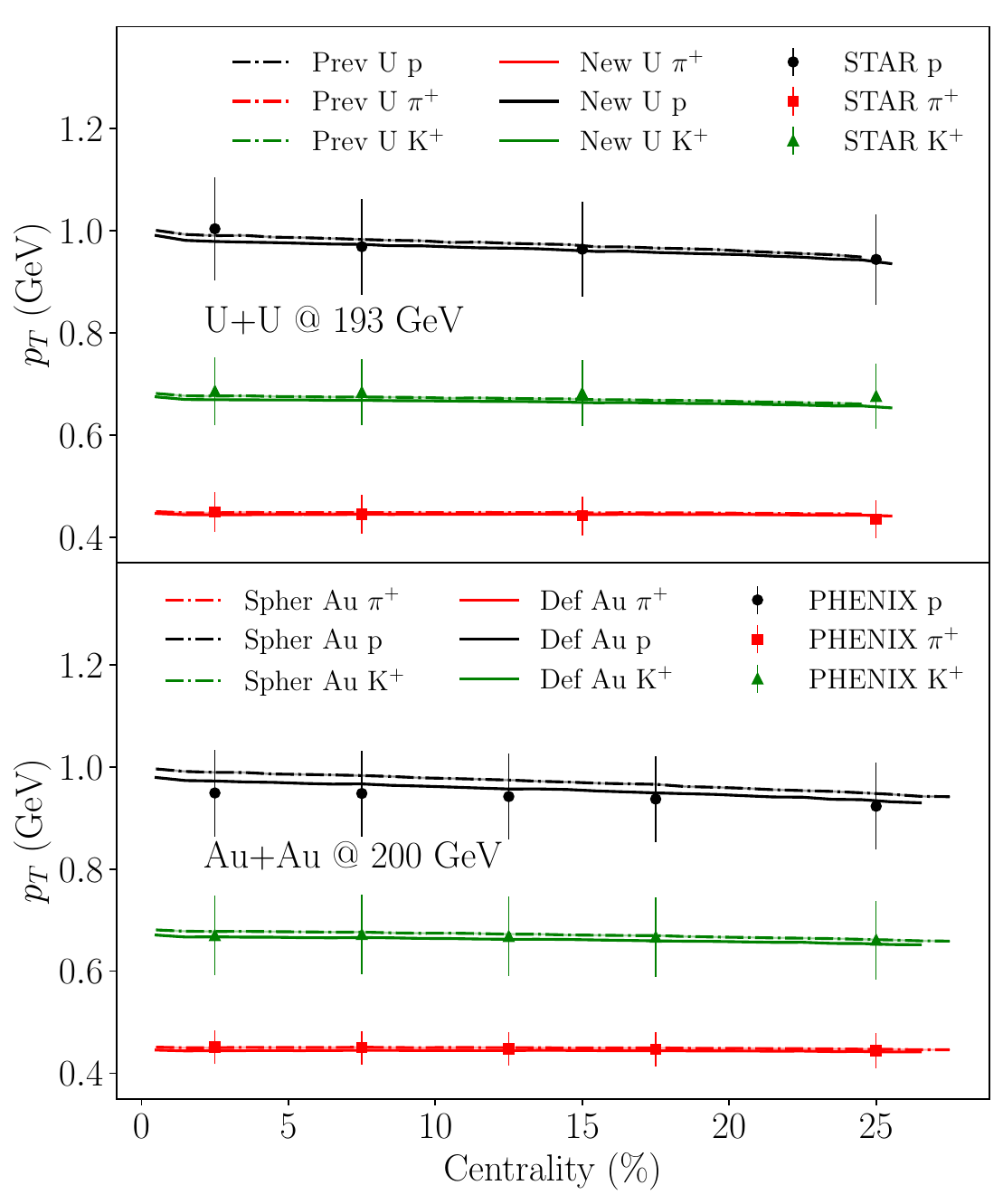}
    \caption{\label{fig:pT} Identified particle mean transverse 
    momentum $\fluc{p_T}$ in $|y| < 0.5$ as a function of centrality in our model. 
    (\textbf{Top}) U configurations compared to results for 
    $\qty{193}{\giga\electronvolt}$ U+U collisions at STAR~\cite{Abdallah_2023}. (\textbf{Bottom}) Au 
    configurations compared to results for 
    $\qty{200}{\giga\electronvolt}$ Au+Au collisions at PHENIX~\cite{Adler_2004}.}
\end{figure}

Fig.~\ref{fig:pT} shows the average transverse momentum of identified particles.
Our model does very well across both systems and all configurations. It does not 
show any obvious preference for any of our parametrizations; the transverse momentum data 
appears insensitive to details of the nuclear deformations. Furthermore, given the relatively
large experimental uncertainties, it would be difficult to argue that specific deformation
effects that may be noticeable in our calculations match experimental data.

\begin{figure}
    \includegraphics[width=\linewidth]{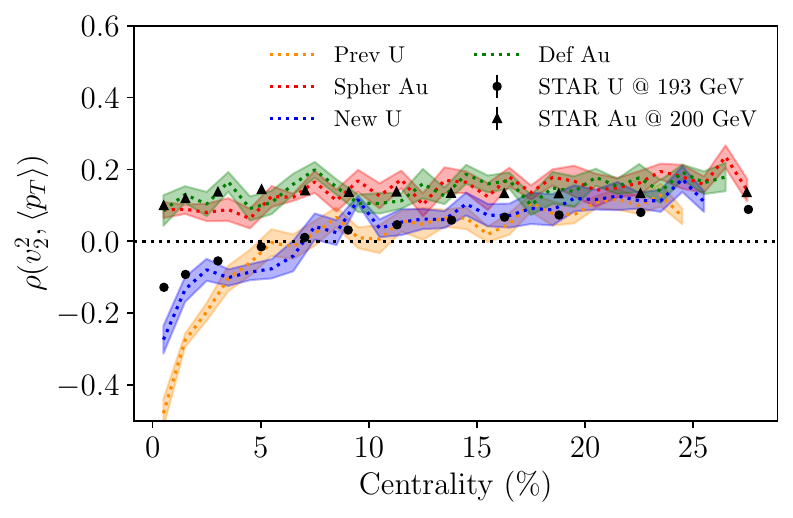}
    \caption{Elliptic flow and 
    $\fluc{p_T}$ correlations as functions of centrality for U+U at $\qty{193}{\giga\electronvolt}$ and Au+Au at $\qty{200}{\giga\electronvolt}$, compared to experimental results from STAR~\cite{starcollaboration2024imaging}. \label{fig:rho}}
\end{figure}

Combining the previous two observables, Fig.~\ref{fig:rho} looks at the correlation 
between the 2-particle cumulant of elliptic flow $v_2\{2\}$ and mean transverse
momentum $\fluc{p_T}$ within a given centrality class. 
We would like to emphasize that our result was posted~\cite{fortier2023comparisonsv1}
{\em before}
the experimental data was made known in~\cite{starcollaboration2024imaging}.
As such, our result is a genuine prediction.

Our model succeeds in reproducing the observable across all 
configurations, with a marked preference for the New U parametrization
(but no clear preference among the Au ones). The main result here, however, is not a preferred 
parametrization, but rather the marked shift from correlation to anti-correlation at 
around $7\%$ centrality. This is a tell-tale sign of large elliptic deformation, 
which our model is capable of reproducing quite elegantly. This shift, again, is due
to the phenomenon described previously: in ultra-central collisions,
body-body collisions will carry more ellipticity and 
relatively lower energy density, while
tip-tip collisions will carry less ellipticity but 
relatively higher energy density.
Therefore, within those centrality classes, elliptic flow $v_2\{2\}$ and mean 
transverse momentum $\fluc{p_T}$ are anti-correlated, with the amount of anti-correlation
increasing as we progress towards the most central collisions. 

This is in stark
contrast to what is observed in collisions of spherically symmetrical nuclei: more 
eccentricity is associated with peripheral collisions and less energy density, thus less 
average transverse momentum. The Au curves, comprised of both a slightly
deformed nucleus and a spherical one, do not exhibit this anti-correlation (nor does
the experimental data). This implies that the observed anti-correlation is due mostly to the 
sizable deformation in the polar angle direction, i.e.~sizable $\beta_2^0$.

\begin{figure}
    \includegraphics[width=\linewidth]{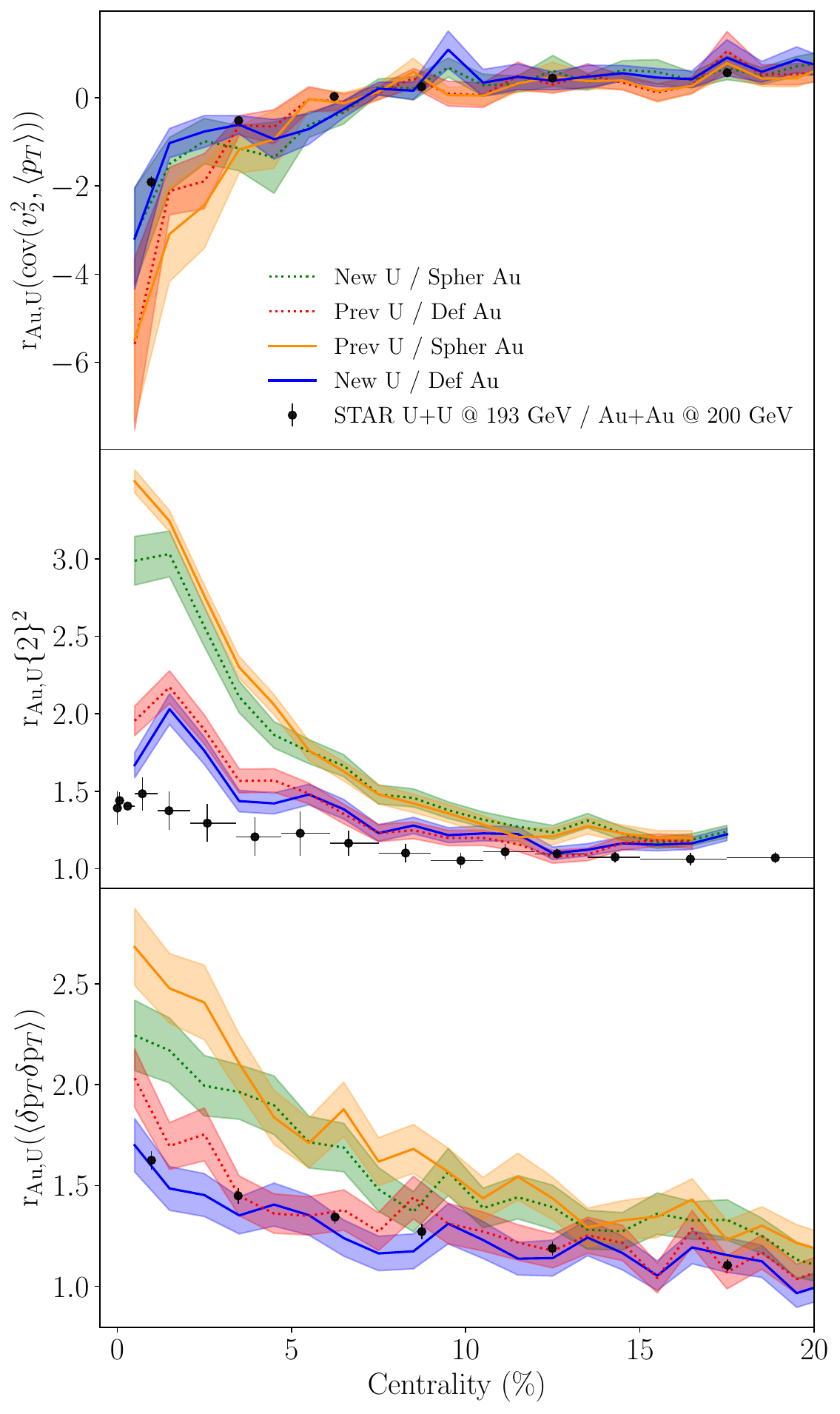}
    \caption{Ratios of (\textbf{top}) elliptic-flow-momentum covariances; (\textbf{middle}) mean squared elliptic flow $v_2\{2\}^2$; (\textbf{bottom}) 2-particle $p_T$ correlator compared to experimental results for U+U and Au+Au collisions at $\qty{193}{\giga\electronvolt}$ and $\qty{200}{\giga\electronvolt}$ respectively at STAR~\cite{starcollaboration2024imaging}.\label{fig:ratios} All observables are defined in detail in our companion paper~\cite{fortier2023comparisons}.}
\end{figure}

Finally, Fig.~\ref{fig:ratios} shows the ratios of observables presented in this letter,
defined as

\begin{gather}
    \textrm{r}_{\rm Au,U}\left(O\right) = \frac{O_{\rm U}}{O_{\rm Au}}
\end{gather}
where $O$ is defined in the figure. These ratios allow for comparisons to experimental data 
which are, in theory, less sensitive to our hydrodynamical evolution~\cite{starcollaboration2024imaging}. It also 
provides an additional angle from which to determine the most appropriate parametrizations.
The ratio of the covariances\footnote{We chose to use the covariances instead of the correlations here in order to be consistent with the provided experimental data. These covariances are the numerators of the whole correlators shown in Fig.~\ref{fig:rho}~\cite{starcollaboration2024imaging,fortier2023comparisons}} show a preference for 
New U and Def Au, although it is only slight. The second ratio is that of the 2-particle cumulant of the elliptic flow, $v_2\{2\}$.
While Fig~\ref{fig:v2} is fairly clear in terms of its parametrization preference,
we opted to include all elliptic flow ratios for completeness.
While the New U and Def Au ratio is closest to the experimental ratio, it overshoots
the experimental data in central collisions. This is mostly due to our 
underestimation of the Au $v_2\{2\}$ in the target range.
This opens the door to future analyses that incorporate more deformed 
Au parametrizations, particularly in the polar angle direction (larger $\beta_2^0$), compared to
the parametrizations used in our calculations, sourced from
~\cite{Ryssens_2023,osti_6477756}. However, as discussed in our assessment of 
Fig.~\ref{fig:rho}, using Au parametrizations with larger $\beta_2^0$ may come at the cost 
of lessening $\rho(v_n\{2\}^2,\fluc{p_T})$, straying away from experimental data. Therefore,
any further analysis will have to carefully balance increasing elliptic flow with preserving
elliptic-flow-momentum correlations.
The last ratio is that of the 2-particle $p_T$ correlators.
This correlator is dominated by short-range
fluctuations, requiring a specific sub-event averaging technique, outlined in our companion
paper~\cite{fortier2023comparisons}, to reproduce its scale.
This ratio once again exhibits sensitivity to initial state geometry, showing
a strong preference for the same pair of parametrizations (New U and Def Au) as 
the other two ratios. Given the ordering of the
ratios, it seems that nuclear deformity correlates directly with higher 2-particle 
$p_T$ correlations in central ($\sim 0-10\%$) collisions. Indeed, ratios using Spher 
Au in the denominator overshoot the experimental ratio by more than $20\%$ across this central window.
Furthermore, the Prev U parametrization, which has higher $\beta_2^0$ than that of 
New U, also overshoots the ratio, albeit in a much narrower window ($\sim 0-4\%$).
The 2-particle $p_T$ correlator was not given its own plot, unlike $v_2\{2\}$ and $\rho(v_n\{2\}^2,\fluc{p_T})$, 
because of its extreme sensitivity to the hydrodynamic medium relative to other 
observables. This plot, along with a detailed description and analysis of this correlator,
can be found in our companion paper~\cite{fortier2023comparisons}.

\prlsection{Summary} We compared calculations using our state-of-the-art, multistage heavy-ion 
collision simulation to a variety of observables which are sensitive to nuclear structure 
and initial state fluctuations. These calculations included four different nuclear 
parametrizations which allowed us to isolate measurements capable of differentiating between 
those differents sets, for our two target systems. While our findings are consistent with the 
the most recent nuclear structure calculations~\cite{Ryssens_2023},
our model suggests that the experimental data involving $^{197}$Au nuclei contains signals 
consistent with a prolate deformation larger than what has been used here, opening the door to 
future analyses.

We provided the first theoretical calculation of the transverse-momentum-flow correlation 
$\rho(v_n\{2\}^2,\fluc{p_T})$ in collisions of deformed nuclei, which proved to mirror 
experimental results. This is 
significant given that our model was never specifically tuned to match this new quantity and 
that it features such a prominent signal of nuclear deformation in the initial state that it 
could potentially be used in future analyses as an indicator of the scale of deformation in 
new collision systems.

Finally, we showed that properly modelling nuclear structure is a key part of 
reproducing a wide range of observables, affirming that any detailed  initial state model 
should include a careful nuclear structure sampling process. These findings mark an exciting development: 
the link between low energy nuclear science with the study of relativistic nuclear collisions, 
highlighting the unity of subatomic physics.

\acknowledgements We would like to acknowledge the support of the entirety of our research group at McGill
University. We also acknowledge insightful conversations with R.~Modarresi Yazdi, 
M.~Heffernan, S.~McDonald, S.~Shi, B.~Schenke C.~Shen, J.~Jia and C.~Zhang. 
This work was funded by the Natural Sciences and Engineering Research Council
of Canada (NSERC) [SAPIN-2018-00024 ; SAPIN-2020-00048]. Cette recherche a \'{e}t\'{e} financ\'{e}e 
par le Conseil de recherches en sciences naturelles et en 
g\'{e}nie du Canada (CRSNG), [SAPIN-2018-00024 ; SAPIN-2020-00048]. Computations were made on the
B\'{e}luga supercomputer system from McGill University, managed by Calcul Qu\'{e}bec
(\url{calculquebec.ca}) and Digital Research Alliance of Canada (\url{alliancecan.ca}).
The operation of this supercomputer is funded by the Canada Foundation for Innovation
(CFI), Minist\`{e}re de l'\'{E}conomie, des Sciences et de l'Innovation du Qu\'{e}bec
(MESI) and le Fonds de recherche du Qu\'{e}bec - Nature et technologies (FRQ-NT).

\bibliography{Bibliography.bib}
\end{document}